\documentclass[twocolumn,preprintnumbers,amsmath,showpacs,amssymb]{revtex4}
\usepackage{epsfig,graphicx}
\usepackage{dcolumn}
\usepackage{bm}
\usepackage{verbatim}

\begin{document}

\title{Waveform sample method of excitable sensory neuron}
\author{Sheng-Jun Wang,$^{1}$ Xin-Jian Xu,$^{2}$ and Ying-Hai Wang$^{1,}$\footnote{For correspondence: yhwang@lzu.edu.cn}}
\address{$^{1}$Institute of Theoretical Physics, Lanzhou University, Lanzhou Gansu 730000, China\\
$^{2}$Department of Electronic Engineering, City University of
Hong Kong, Kowloon, Hong Kong, China}

\date\today

\begin{abstract}
We present a new interpretation for encoding information of the
period of input signals into spike-trains in individual sensory
neuronal systems. The spike-train could be described as the
waveform sample of the input signal which locks sample points to
wave crests with randomness. Based on simulations of the
Hodgkin-Huxley (HH) neuron responding to periodic inputs, we
demonstrate that the random sampling is a proper encoding method
in medium frequency region since power spectra of the
reconstructed spike-trains are identical to that of neural
signals.
\end{abstract}


\maketitle

In the last decade the phenomenon of the stochastic resonance of
excitable neuron has been extensively studied both experimentally
\cite{Douglass, Bulsara, Levin, Braun} and theoretically
\cite{Wiesenfeld, longtin, Moss}. In these studies stochastic
excitable sensory neurons or neuron models subject to periodic
signals. Stochastic resonance means that the ability of neurons
for responding to the weak signal is enhanced by noise. The power
spectral density of action potential was used as one of common
methods for estimating the sensitivity of stochastic neurons to
weak input signal \cite{Gammaitoni}. The power spectrum revealed
the frequency information of input signal from the time series
action potential \cite{Douglass, Verechtchaguina}. However, the
code for encoding and decoding in neuron is not clarified at the
moment \cite{Rieke}. A approximate theory for modelling neuron
firing in the presence of noise and a periodic stimulus has been
proposed \cite{Wiesenfeld}, and the process of neuron conveying
information was described by inhomogeneous Poisson point process.
In this paper, we didn't treat the spike-train as a poisson
process. Our source of inspiration is based on the phase locking
\cite{Bear}. We investigated the correlation exhibited in time
axis between the temporal sequences of firing events and the input
signals, and proposed the waveform sample method used by neurons
for processing periodic signals.

We made a study based on a popular Hodgkin-Huxley (HH) neuron
model which was originally proposed to account for the property of
squid giant axons \cite{Hodgkin} and has been generalized with
modifications of ion conductances \cite{Koch}. It describes the
spiking behavior and refractory properties of real neurons and
serves as a paradigm for spiking neurons based on nonlinear
conductance of ion channels \cite{Hodgkin}. The dynamics of HH
model is described by the following set of coupling ordinary
differential equations: one for the membrane potential $V$ and the
other three for the gating variables: $m$, $n$, and $h$; that is,
\begin{align}
&\dot{V}=(I_{ion}(t)+I_{ext}(t)+I_{noise})/C_m, \tag{1a}\label{eq1a}\\
&\dot{m}=(m_{\infty}(V)-m)/\tau_{m}(V), \tag{1b}\label{eq1b}\\
&\dot{h}=(h_{\infty}(V)-h)/\tau_{h}(V), \tag{1c}\label{eq1c}\\
&\dot{n}=(h_{\infty}(V)-n)/\tau_{n}(V), \tag{1d}\label{eq1d}
\end{align}
where
\begin{equation}
I_{ion}(t)=-g_{na}m^{3}h(V-V_{na})-g_{k}n^{4}(V-V_{k})-g_{l}(V-V_{l}).
\tag{1e}\label{eq1e}
\end{equation}
The ionic current $I_{ion}(t)$ describes the ionic transport
through the membrane, and includes the sodium ($I_{na}$),
potassium ($I_k$), and leak ($I_l$) currents. The constants
$g_{na}$, $g_k$, and $g_l$ are the maximal conductances for ion
and leakage channels, and $V_{na}$, $V_k$, $V_l$ are the
corresponding reversal potentials; $m_\infty$, $h_\infty$,
$n_\infty$ and $\tau_m$, $\tau_h$, $\tau_n$ represent the
saturation values and relaxation times of the gating variables
\cite{Yuyuguo}. Detailed values of parameters can be found in
\cite{Hodgkin, Hansel}. $I_{ext}$ is the external stimulus
received by neuron. We model the noisy current as an exponentially
correlated colored noise
\begin{equation}
\tau_d\frac{dI_{noise}}{dt}=-I_{noise}+\sqrt{2D}\xi(t),
\tag{2}\label{eq2}
\end{equation}
where $\xi(t)$ is Gaussian white noise, $D$ and $\tau_d$ is the
intensity and the correlation time of the colored noise,
respectively. In the following numerical simulations we took
$\tau_d=2$msec. Differential equations given by Eqs.
(\ref{eq1a})-(\ref{eq1e}) were integrated by the forth-order
Runge-Kutta method. The colored noise in Eq. (\ref{eq2}) was
solved following ideas by Fox \emph{et al.} \cite{Fox}. The time
step was $\triangle t=0.02$ msec.

We use spike-train rather than membrane potential to represent the
temporal sequence of firing events of neurons. Because it is
believed that most of information in neural systems is coded in
the time sequence of action potential \cite{Moore}. The
spike-train is a binary time series with a value $1$ at the time
of action potential generations and $0$ at other times. The spikes
were defined as 20mV-level crossing from below to above values of
the membrane potential $V$.

\begin{figure}
\centerline{\epsfxsize=9cm \epsffile{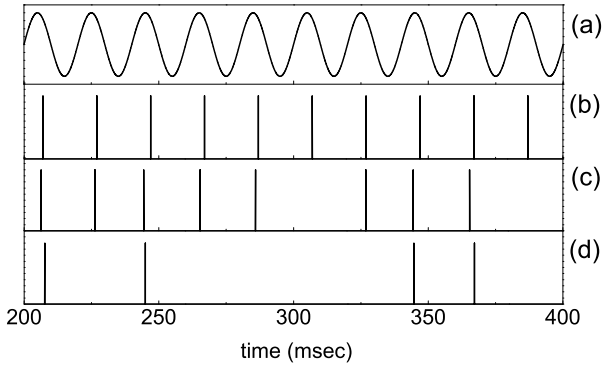}}

\caption{The waveform of the input periodic signal and three
simulated spike-trains from the HH neuron. (a) The waveform of the
signal used in simulations. The coordinates of signal was ignored.
(b) The spike-train obtained by inputting strong signal
($A_1=2.0\mu A/cm^2$) into the HH neuron and noise absent. (c) The
spike-train produced by added noise ($D=5$) to the strong sine
signal used to obtain (b). (d) The spike-train recorded from
simulations using the sum of the weak sinusoidal signal
($A_1=1.0\mu A/cm^2$) and noise ($D=5.0$) as stimulus.}
\label{fig1}
\end{figure}

\begin{figure}
\centerline{\epsfxsize=9cm \epsffile{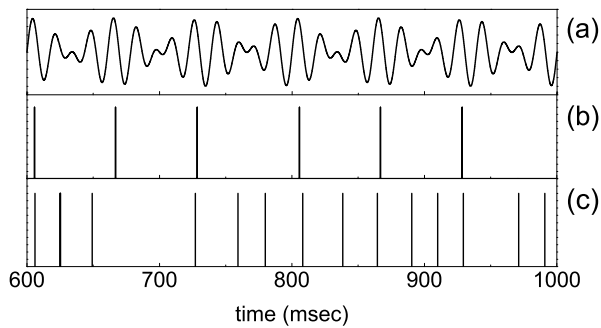}}

\caption{(a) The waveform of the signal which is the sum of two
sine wave which have frequency $f_1=0.05kHz$ and $f_2=0.065kHz$,
respectively. (b) The spike-train produced by the HH neuron which
received the combined signal without noise. (c) The spike-train
recorded from the stochastic HH neuron ($D=10.0$) responding to
the combined signal ($A_2=1.0\mu A/cm^2$). }\label{fig2}
\end{figure}

Some examples of spike-trains produced by the computer simulations
of HH neurons responding to periodic signal with or without noise
were presented. First, we considered a pure tone sinusoidal
external stimulus:
\begin{equation}
I_{ext}=A_1sin(2 \pi f t). \tag{3}\label{Ip1}
\end{equation}
In Fig. \ref{fig1} we plotted the waveform of the pure tone signal
with frequency $f=0.05kHz$ followed by three spike-trains recorded
from three trials with different conditions. The curve in Fig.
\ref{fig1}(a) represented precisely the phase of sinusoidal
signals used in different trials, but intensity was not included.
When the sine signal is strong, $A_1=2.0\mu A/cm^2$, and noise is
absent, the spike-trains in Fig. \ref{fig1}(b) was activated.
There is a spike in each period of signal. And the spikes fire
precisely at a specific phase of sine signal during every positive
half-cycle. There is a common delay between spikes and
corresponding wave crests, which does not affect the temporal
structure of the spike-train and the correlation between the input
signal and the spike-train. According to the phase locking of
firing to sinusoidal signal, we simplified the description of the
correlation between the input signal and the spike-train as that
every wave crest of the pure tone sine wave activates a spike. In
the second trial, we added noise into the HH neuron which received
the sinusoidal signal used upper. The intensity of noisy current
was $D=5.0$. The spike-train recorded from the simulation was
shown in Fig. \ref{fig1}(c). Notice that most of the spikes
deviated from the positions contrasting to that in the above
spike-train and few spikes disappeared randomly. The size of
deviation is random and small. In the third trial, we decrease the
amplitude of periodic signal to $A_1=1.0\mu A/cm^2$, and remained
the intensity of noise as $D=5.0$. The spike-train computed from
the trial was illustrated by Fig. \ref{fig1}(d). Under this
condition, the number of spikes had a sharp decrease compared with
the spike-train activated by strong signal shown in Fig.
\ref{fig1}(c). It is random that whether a wave crest activates a
spike. The spikes in the spike-train also deviate from the
positions of the wave crests. Although randomness entered the
spike-train when noise was presence, the phase locking was
affected slightly by noise.

Pure tone stimulations have been used in many studies of neuronal
response \cite{Gammaitoni}. Subsequently, we considered the signal
which is a sum of two sinusoidal functions with different
frequencies:
\begin{equation}
I_{ext}=A_2(sin(2 \pi f_1 t)+sin(2 \pi f_2 t)). \tag{4}\label{Ip2}
\end{equation}
We took the frequencies of two sinusoidal functions as
$f_1=0.05kHz$, $f_2=0.065kHz$. In Fig. \ref{fig2} the waveform of
the combined signal and the spike-trains activated by it were
illustrated. In Fig. \ref{fig2}(b), during a period of the input
signal, from 600 msec to 800 msec, three spikes fired and locked
to the three highest wave crests. When noise was added to the
neuron, the number of spikes increased. Some spikes fired randomly
at the small neighborhoods of weaker wave crests. The stronger a
positive half-wave is, the larger probability it gives rise to a
spike. According to the correlations between the input signal and
the spike-train, we propose an interpretation of the encoding
mechanism. The phase locking phenomena suggest that the response
to periodic signal can be considered as a waveform sample: when a
sensory neuron receives the periodic signal, it samples the input
signal through the nonlinear response and the signal was sampled
as the action potential. The sampling method has the following
properties. (\romannumeral1) Sample points appear in the
neighborhoods of the wave crests of input signals.
(\romannumeral2) It is unnecessary that every wave crest is
sampled. The probability that a wave crest is sampled depends on
the strength of it. (\romannumeral3) The positions of the sampling
points distribute randomly in the neighborhoods of the wave crest
of signals. (\romannumeral4) The values of the sampling points do
not depend on the signals which are sampled. All samples have the
same value.

To demonstrate the waveform sample method, we used it to recover
the basic characteristics of neural signals. The power spectra of
the spike-train was used to evaluate the frequency information of
neural signals and assess the coherence of the spiking activity
with the signal frequency, which has been used extensively in the
studies of stochastic resonance as mentioned above. We computed
the power spectral density from simulating the process that an
individual HH neuron responded to periodic signals and the
Monte-Carlo (MC) simulations of the waveform sample method. The
power spectra were computed from 200 averages of the power spectra
for the time series of the spike-train with length $2^{17}$ using
the fast Fourier transform \cite{press}.

\begin{figure}
\centerline{\epsfxsize=9cm \epsffile{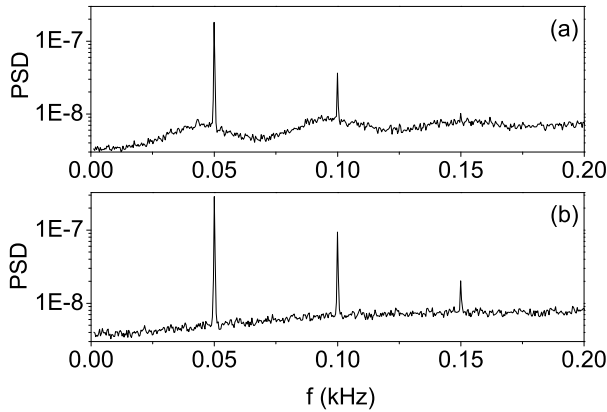}}

\caption{(a) The power spectrum of the spike-train excited by the
weak pure sinusoidal current ($A=1.0\mu A/cm^2$) with noise
($D=5.0$). (b) The power spectrum of the reconstructed spike-train
produced by the MC simulation of the waveform sample method with
$I_c=2.0\mu A/cm^2$ and $\sigma=2.0$msec. }\label{fig3}
\end{figure}

Let us first consider the scheme that a stochastic HH neuron
receives the pure tone sinusoidal signal described by Eq.
(\ref{Ip1}). We took the amplitude and frequency of signal as
$A_1=1.0 \mu A/cm^2$ and $f=0.050 $ kHz. This signal is too weak
to activate HH neuron without noise (the critical amplitude is
$A_c=1.52 \mu A/cm^2$ when $f=0.050k$Hz). We calculated the power
spectrum of the spike-train in the presence of noise ($D=5.0$).
The power spectrum was shown in Fig. \ref{fig3}(a). In the power
spectrum, the main feature due to the periodic signal is the
narrow peak at the fundamental frequency riding on a broad noise
background. Additionally, the harmonic of the fundamental
frequency exists in the power spectrum. This result agrees
excellently with the experiment in Ref. \cite{Douglass}. The power
spectrum quantify the information content of the spike-train
without invoking any specific encoding mechanism. The peak of
power spectral density located at the fundamental frequency proves
that the information of the period of the signal must be encoded
in the spike-train by the individual HH sensory neuron. As an
example, we consider a simple rule of the waveform sample. The
probability that a wave crest gives a sample point is proportional
to the amplitude of signal:
\begin{equation}
P_i=I_i/I_c, \tag{5}\label{pro1}
\end{equation}
where, $I_i=A_1$ is the amplitude of the pure tone sinusoidal
signal, $I_c$ is constant which is larger than $A_1$. The
positions of sample points were subjected to Gaussion
distribution:
\begin{equation}
p(s_i)=\frac{1}{\sqrt{2\pi}\sigma}\exp{(-\frac{(s_i-t_i)^2}{2\sigma^2})},
\tag{6}\label{dis}
\end{equation}
where $t_i$ is the position of the $i$th wave crest, $s_i$ is the
position of sample point that locks in the neighborhood of $t_i$.
We made MC simulations using the rule described by Eqs.
(\ref{pro1}) and (\ref{dis}) with $I_c=2.0\mu A/cm^2$ and
$\sigma=2.0$msec. The power spectrum obtained from the MC
simulation was compared with that of the HH neuron in Fig.
\ref{fig3}. Although the background noise in the power spectra are
different, the frequency structure of HH neuron has been recovered
virtually. The difference between background noise in Fig.
\ref{fig3}(b) and that in Fig. \ref{fig3}(a) is due to the simple
rule of waveform sample which does not involve the dynamics of HH
neuron.

\begin{figure}
\centerline{\epsfxsize=9cm \epsffile{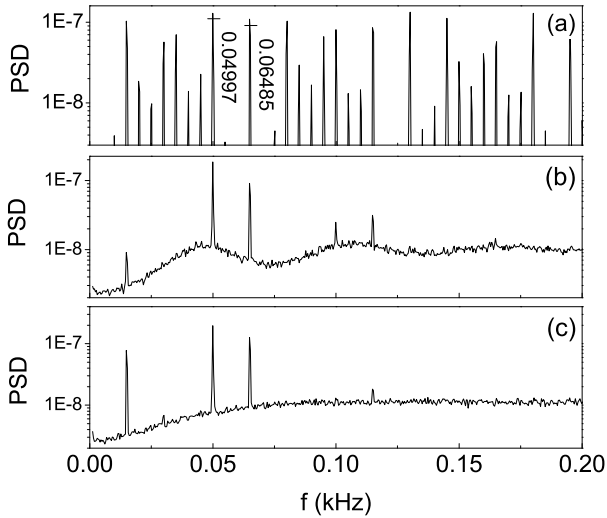}}

\caption{(a) The power spectrum recorded from the simulation of HH
neuron responding to the periodic signal described by Eq.
(\ref{Ip2}) without noise. The parameter were taken as
$f_1=0.05kHz$, $f_2=0.065kHz$ and $A_2=1.0\mu A/cm^2$. (b) The
power spectrum produced by stochastic HH neuron ($D=10.0$)
responding to the combined signal. (c) The power spectrum obtained
from the MC simulation of waveform sample.}\label{fig4}
\end{figure}

Then we consider the combined signal described by Eq. (\ref{Ip2}).
The parameters were taken as $A_2=1.0 \mu A/cm^2$, $f_1=0.05kHz$,
and $f_2=0.065kHz$ in simulations. Under this condition, the HH
neuron can be activated by the combined current without the
assistance of noise. In the power spectrum, as illustrated by Fig.
\ref{fig4}(a), the fundamental frequencies give two high and
narrow peaks which were marked by the labels using their positions
in frequency domain. Also lots of additional frequency components
arise and some of which are as strong as fundamental frequencies.
The frequency structure exhibited in this power spectrum was given
by the precisely phase locking spike-train shown above in Fig.
\ref{fig2}(b). When noise was present, we simulated the process
that HH neuron responds to the combined signal. In the case of
noise intensity was taken as $D=10.0$, the power spectrum was
computed and shown in Fig. \ref{fig4}(b). The frequency structure
has a observably change compared with Fig. \ref{fig4}(a). The
fundamental frequencies $f_1$ and $f_2$ give the highest peaks in
the power spectrum. Most of the additional frequency components
disappear. The surviving additional components locate at
$f_2-f_1$, $2f_1$ and $f_1+f_2$ with weak strength. Noise affects
the frequency structure of the output spike-train of sensory HH
neuron and plays an optimized role in distinguishing the
fundamental frequency components in the power spectrum. In the MC
simulation of the waveform sample of the combined periodic signal,
the sample rule described by Eqs. (\ref{pro1}) and (\ref{dis}) was
used again. The symbol $I_i$ was the height of the $i$th wave
crest of the combined periodic signal. We took the parameters as
$I_c=2.0\mu A/cm^2$ and $\sigma=3.0$msec. The power spectrum of
the waveform sample was shown in Fig. \ref{fig4}(c), and it got
the primary characters of the frequency structure of the Fig.
\ref{fig4}(b). The result demonstrates that the randomness in the
waveform sample method captures the effect of noise in sensory
neuron. The sample rule used in the MC simulation was too simple
and crude to depict the neural process exactly, but the main
characteristics of spike-trains of the HH model have been obtained
from the waveform sample. The waveform sample method captures the
nature of the encoding mechanism that sensory neuron responds to
periodic signals.

The present work is the further investigation of the stochastic
process theory of firing events \cite{Wiesenfeld}, where the
frequency structure of power spectra were described elegantly. We
obtained the subtler correlation between the spike-train and the
input signal, and proposed a waveform sample method as an
intuitional interpretation of the encoding mechanism. The waveform
sample method is different from the conventional sample method
used for processing signals, such as the even interval sample used
for discrete Fourier transform \cite{Brigham} which does not give
rise to the additional frequency components in power spectra. If
the frequency of the signal received by neuron is low, the neuron
will fire more than once in a period of signal. On the other hand,
the dynamics of HH neuron can not lock the spike to every wave
crest of the signal in the high frequency region because of the
refractory properties. Although it is not a proper encoding method
of neurons in low and high frequency regions, it can also convey
the frequency information into the sample points in all these
frequency regions. Additionally, the waveform sample method
suggests a method for reconstructing the spike-train and proposes
a chance to utilize the neural encoded mode without invoking
neuronal models.

\end{document}